\documentclass[twocolumn]{revtex4}[12pt]
\usepackage{graphicx}
\begin{document}
\title{Nucleus-acoustic solitary waves and double layers in a magnetized degenerate quantum plasma}
\author{B. Hosen$^{1*}$, M. G. Shah$^{2}$, M. R. Hossen$^{3}$, and A. A. Mamun$^{1}$}
\address{$^{1}$Department of Physics, Jahangirnagar University, Savar, Dhaka-1342,
Bangladesh\\ $^{2}$ Department of Physics, Hajee Mohammad Danesh
Science and Technology University, Dinajpur-5200, Bangladesh\\
$^{3}$Department of General Educational Development, Daffodil
International University, Dhanmondi, Dhaka-1207, Bangladesh\\
Email$^{*}$: hosen.plasma@gmail.com}

\begin{abstract}
The properties of nucleus-acoustic (NA) solitary waves (SWs) and
double layers (DLs) in a four-component magnetized degenerate
quantum plasma system (containing non-degenerate inertial light
nuclei, both non-relativistically and ultra-relativistically
degenerate electrons and positrons, and immobile heavy nuclei)
are theoretically investigated by the reductive perturbation
method. The Korteweg-de Vries (K-dV), the modified K-dV (MK-dV),
and the Gardner equations are derived  to examine the basic
features (viz. amplitude, speed, and width) of NA SWs and DLs. It
is found that the effects of the ultra-relativistically degenerate
electrons and positrons, stationary heavy nuclei, external
magnetic field (obliqueness), etc. significantly modify the basic
features of the NA SWs and DLs. The basic features and the
underlying physics of NA SWs and DLs, which are relevant to some
astrophysical compact objects including white dwarfs and neutron
stars, are pinpointed.
\end{abstract}

\maketitle
\section{Introduction}
%%%%%%%%%%%%%%%%%%%%%%%%%%%%%%%%%%%%%%%%%%%%%%%%%%%%%%%%%%%%%%%%%%%%%%%%%%%%%%%%%%%%%%
There has been a great deal of interest in understanding the
plasma fluid dynamics inside the astrophysical dense
electron-positron-ion (EPI) plasmas. The EPI plasma is an
omnipresent ingredient of many astrophysical systems or regions,
such as early universe \cite{Misner1973}, active galactic nuclei
\cite{Begelman1984,Miller1987,Tribeche2009}, pulsar magnetosphere
\cite{Liang1998,Michel1982}, white dwarfs
\cite{Hossen2014a,Hossen2014b}, polar regions of neutron stars
\cite{Michel1991}, solar atmospheres
\cite{Goldreich1969,Hansen1988}, inner regions of the accretion
disc surrounding black holes \cite{Rees1971}, center of our
galaxy \cite{Burns1983}, etc. A number of invesitgations have
been made on small amplitude solitons in EPI plasmas in the
presence of significant percentage of positrons \cite{Popel1995}.
The annihilation time of positrons in the plasma is long compared
with typical particle confinement times. However, for long
lifetime of positrons, different types of  linear waves and
associated nonlinear structures can be generated  in such EPI
plasmas \cite{Tandberg-Hansen1988,Piran1999}. The characteristics
of linear waves and  nonlinear structures have been influenced
considerably due to the presence of positron component along with
electrons and ions in most space plasmas.

It is now well established that the effect magnetic field
significantly modifies the basic properties of linear and
nonlinear waves in EPI plasmas  which are, in fact, magnetized for
most laboratory, space, and astrophysical situations
\cite{Matthaeus2005,Jehan2008,Bains2010,Mahmood2003,Sabry2012}.
The electromagnetic forces and degenerate pressures of electrons
and positrons give rise to different types of fluctuations at very
short length and time scales. The effect of magnetic field on the
stability behaviour of low-frequency electrostatic IA waves in
EPI plasmas is examined by Jehan \textit{et al.}
\cite{Jehan2008}. Bains \textit{et al.} \cite{Bains2010} studied
the modulational instability of IA waves encompasses in
magnetized quantum EPI plasmas. The large amplitude IA waves
propagation in magnetized EPI plasma using Sagdeev potential
approach is studied by Mahmood \textit{et al.}
\cite{Mahmood2003}.  Sabry \textit{et al.} \cite{Sabry2012}  have
taken the typical magnetized EPI parameters, which are  relevent
to white dwarfs and corona of magnetars, to examine the IA freak
waves in ultra-relativistic degenerate EPI plasmas.

Recently, there has been a renewed interest in studying the
relativistic degenerate dense plasmas due to its existence in
interstellar compact objects, such as white dwarfs
\cite{Zobaer2012,Akhter2013,mr2014a,mr2014b}, neutron stars
\cite{Shapiro1983,Mamun2010a,Mamun2010b} and in tense laser
plasma experiments \cite{Marklund2006,Mourou2006}. It is observed
that at a very large number densities, when the electron and
positron Fermi energies become dominant over the electron and
positron thermal energies, then the electron-positron thermal
pressure can be ignored in comparison with the electron-positron
degeneracy pressures. Within a astrophysical objects, the lower
energy state is filled of electrons so additional electrons be
cannot given up energy to the lower energy state and therefore
they generate degeneracy pressure. In case of astrophysical
compact object, like white dwarf, the average density could be
varied from $10^{6}$ $gmcm^{-3}$ to $10^{8}$ $gmcm^{-3}$, the
degenerate electron number can be of the order of $10^{30}$
$cm^{-3}$, and the average inter-particle distance is of the order
of $10^{-10}$ $cm$ which is of the order of $10^{-13}$ $cm$ for
neutron star \cite{Shapiro1983,Mamun2010a,Koester1990,Azam2005}.
In such compact objects the light nuclei can be assumed to be
inertial, whereas the electrons and positrons are taken to follow
the degeneracy pressure in order to support these objects against
the gravitational collapse. It is notable that the basic
constituents of white dwarfs are mainly positively and negatively
charged heavy elements like carbon, oxygen, helium with an
envelope of hydrogen gas. The existence of heavy elements
(positively and negatively) is found to form in a prestellar
stage of the evolution of the universe, when all matter was
compressed to extremely high densities. The average number
density of heavy particles is of the order of $10^{29}$ $cm^{-3}$
where distance between heavy particles is of the order of
$10^{-10}$ $cm$ (for white dwarfs) \cite{Koester1990}.

A general expression for the degenerate pressure of relativistic
ions and electrons is given by Chandrasekhar. The equation of
state regarding the degeneracy of plasma species is presented at
earlier in case of both the non-relativistic
$({P_f}\propto{n_f}^{5/3})$ and the ultra-relativistic
$({P_f}\propto{n_f}^{4/3})$  limits by Chandrasekhar
\cite{Chandrasekhar1931,Chandrasekhar1935}, where $P_f$ shows the
degenerate pressure and $n_f$ describes the number density of
plasma species $f$, respectively. To demonstrate the equation of
state, Chandrasekhar introduced that for the nonrelativistic
degenerate electrons,
$\gamma=\frac{5}{3};~K=\frac{3}{5}\left(\frac{\pi}{3}\right)^{\frac{1}{3}}
\frac{\pi\hbar^2}{m}\simeq\frac{3}{5}\Lambda_c\hbar c$ where $K$
is the proportionality constant, $\Lambda_c=\pi \hbar/mc=1.2\times
10^{-10}~cm$, and $\hbar$ is the Planck constant divided by
$2\pi$ and for the ultrarelativistic degenerate electrons,
$\gamma=\frac{4}{3};~K=\frac{3}{4}\left(\frac{\pi^2}{9}\right)^{\frac{1}{3}}
\hbar c\simeq\frac{3}{4}\hbar c$. Using the Chandrasekhar limit,
Taibany \textit{et al.} \cite{El-Taibany2012} considered an ultra
dense plasma to study the properties of electromagnetic
perturbation modes (fast and slow modes) and clarified that the
roles of degeneracy of electrons and positrons, enthalpy
corrections, strength of magnetic field, and relativistic factor
greatly modify the magnetosonic perturbation modes. Pakzad and
Javidan \cite{Pakzad2011} studied the IASWs in dissipative
plasmas containing relativitic ions, nonthermal electrons, and
maxwellian positrons. Ting \textit{et al.} \cite{Ting1992}
thought out a relativistic plasma system with nonisothermal
electrons and examined the transient features of solitary waves.
Chatterjee \textit{et al.} \cite{Chatterjee2012} analyzed the
nonlinear propagation of IASWs in an unmagmetized plasma
comprising of nonthermal electrons and positrons, and singly
charged adiabatically hot positive ions in planar ana nonplanar
geometries. Moslem \textit{et al.} \cite{Moslem2007} studied the
two-dimensional non-linear acoustic excitations in EPI plasmas
that are applicable to the accretion discs of active galactic
nuclei, where the ion temperature are 3-300 times higher than
those of electrons. Haque and Saleem \cite{Haque2003} studied
large amplitude two-dimensional IA and drift wave vortices in
magnetized EPI plasmas, where the electrons and positrons were
assumed to be Boltzmann-distributed. Masood and Mushtaq
\cite{Masood2008} investigated the linear properties of obliquely
propagating magnetoacoustic waves in EPI quantum magnetoplasmas
and found that the corrections significantly modify the
propagation of fast and slow magnetoacoustic waves in both
plasmas. Rizzato \cite{Rizzato1988} studied the localization of
weakly nonlinear circularly polarized electromagnetic waves in a
cold plasma made up of electrons, positrons, and ions. Shah
\textit{et al.} \cite{Shah2015a,Shah2015b,Shah2015c,Shah2015d}
studied an unmagnetized degenerate quantum plasma and
investigated the effects of relativistic degenerate electrons and
positrons and plasma particle number densities on the propagation
of positron-acoustic solitary or shock waves. Shuchy \textit{et
al.} \cite{Shuchy2012} studied the electron-acoustic (EA) and
dust-electron-acoustic Gardner Solitons (GSs) and DLs with
two-temperature-electron (hot and cold) following Boltzmann
distributions. Wang \textit{et al.} \cite{Wang2009} investigated
the effect of an external uniform magnetic field on the
propagation of  solitary waves in a in the weakly relativistic
magnetized multi-ion plasma containing electrons and light and
heavy ions. Labany and Shaaban \cite{Labany1995} have considered
nonlinear IA waves in a weakly relativistic plasma consisting of
warm ion-fluid with non-isothermal electrons. The propagation of
electrostatic and electro-magnatic waves in quantum plasma with
relativistic degenerate electrons has been analyzed by Khan
\cite{Khan2012}. Mamun and Shukla \cite{Mamun2010a} considered
the reductive perturbation technique to investigate the solitary
waves in ultra-relativistic degenerate dense plasmas. Zeba
\textit{et al.} \cite{Zeba2012} studied the nonlinear IA waves in
the dense unmagnetized EPI plasmas with ultra-relativistic
degenerate electrons and positrons. Masood \textit{et al.}
\cite{Masood2010} studied the electro-magnetic wave equation for
relativistic degenerate quantum magnetized plasmas. More
recently, Hossen \textit{et al.}
\cite{rasel2014a,rasel2014b,rasel2014c,rasel2015} investigated the
basic features of different nonlinear acoustic waves in the
presence of heavy elements in a relativistic degenerate plasma
system. But this attempts is valid only for unmagnetized case.
However, to the best of knowledge, no theoretical investigation
has been made on the nonlinear properties of NA SWs in magnetized
degenerate quantum plasmas with relativistic degenerate electrons
and positrons. Therefore, in our present work, we attempt to study
the basic features of NA SWs by deriving the K-dV, MK-dV and
Gardner equations in magnetized degenerate quantum plasma
containing non-degenerate inertial light nuclei, both
non-relativistically and ultra-relativistically degenerate
electrons and positrons, and immobile heavy nuclei.

The manuscript is organized as follow. The theoretical model
describing the dynamics of the NA waves is described in Sec. II.
The nonlinear dynamical equation, namely K-dV (MK-dV) Gardner
equations and its solitay wave solution are derived and
interpreted in Sec. III (IV) V. A brief discussion is finally
provided in Sec. VI.

%%%%%%%%%%%%%%%%%%%%%%%%%%%%%%%%%%%%%%%%%%%%%%%%%%%%%%%%%%%%%%%%%%%%%%%%%%%%%%%%%%%%%%%%%%%%%%%%%%%%%%%%%%%%%%
\section{Theoretical Model}
We consider a four component magnetized quantum plasma system
consisting of non-degenerate inertial light nuclei, both
non-relativistic and ultra-relativistic degenerate electrons and
positrons, and immobile heavy nuclei. At equilibrium,
$n_{e0}=n_{i0}+n_{p0}+Z_hn_{h0}$, where $n_{e0}$, $n_{i0}$,
$n_{p0}$ are the unperturbed number densities of degenerate
electron, inertial light nuclei, degenerate positron,
respectively and $Z_h$ is the number of negative ions residing
onto the heavy nuclei surface. The positively charged static
heavy nuclei participate only in maintaining the quasi-neutrality
condition at equilibrium. The mass of the electrons and positrons
can be considered inertialess in comparison to light nuclei mass,
and one can describe the degenerate pressure through Eqs. (1) and
(2). Generally, in a relativistically degenerate plasma system,
pressure arises due to the combined effect of Pauli's exclusion
principle and Heisenberg's uncertainty principle, and depends
only on the number density of constituent particles, but not on
its thermal temperatures \cite{Mamun2010a,Koester1990}. The
electron and positron inertia can in fact be neglected if we
consider that NA electrostatic waves move at phase velocity
($V_p$) which is much higher than the light nuclei thermal speed
and yet in turn much lower than the electron and positron thermal
speed: $V_{th,i}<<V_p<<V_{th,e,p}$. The dynamics of nonlinear NA
waves in the presence of the external magnetic field
$\textbf{B}=\hat{z}B_0$ is governed by the following momentum
equation
\begin{eqnarray}
&&\hspace*{-10mm}\nabla\phi-\frac{K_1}{n_{e}}\nabla{n_e}^\gamma=0,\label{B2a}\\
&&\hspace*{-10mm}\nabla\phi-\frac{K_2}{n_{p}}\nabla{n_p}^\gamma=0,\label{B2b}
\end{eqnarray}
and the non-degenerate inertial light nuclei equations composed
of the light nuclei continuity and light nuclei momentum
equations are given by
\begin{eqnarray}
&&\frac{\partial n_i}{\partial t} + \nabla
.({n_i}{\textbf{u}_i})= 0,
\label{B2c}\\
&&\frac{\partial \textbf{u}_i}{\partial t} +
({\textbf{u}_i}.\nabla)u_i=- \nabla \phi+
\omega_{ci}({\textbf{u}_i}\times{\hat{z}}),\label{B2d}
\end{eqnarray}
The equation that is closed by Poisson's equation
\begin{eqnarray}
&&\nabla^2
\phi={n_e}{\alpha_e}-n_i-{n_p}{\alpha_p}-{Z_h}\mu,\label{B2e}
\end{eqnarray}
where $n_s$ is the perturbed number densities of species s (here
s = i, e, p inertial light nuclei, degenerate electron,
degenerate positron, respectively). $\textbf{u}_i$ is the light
nuclei fluid speed normalized by NA speed
$C_{i}=(m_{e0}c^2/m_{i0})^{1/2}$ with $m_{e0}$ ($m_{i0}$) being
the electron (light nuclei) rest mass, $c$ is the speed of light
in vacuum, $\phi$ is the electrostatic wave potential normalized
by $m_{e0}c^2/e$. Here $\mu\ = (n_{h0}/n_{i0})$ is the heavy
nuclei to light nuclei number density ratio, $\alpha_e =
(n_{e0}/n_{i0})$ is the electron to light nuclei number density
ratio and $\alpha_p = (n_{p0}/n_{i0})$ is the positron to light
nuclei number density ratio. The nonlinear propagation of usual
NA waves in EPI plasma can be recovered by setting $\mu=0$. The
time variable ($t$) is normalized by ${\omega_{pi}}=(4 \pi
n_{i0}e^2/m_{i0})^{1/2}$, and the space variable ($x$) is
normalized by $\lambda_{s}=(m_{e0}c^2/4 \pi n_{i0}e^2)^{1/2}$.
The numerical value of $\lambda_s$ is of the order of $10^{-10}
cm$ \cite{Mamun2010a}. We have defined the parameters appeared in
Eq. (1) as relativistic factor for degenerate electrons,
$K_1=n_{e0}^{\gamma-1}K_e/{m_{e0}}{c}^2$ and in Eq. (2) as
relativistic factor for degenerate positrons,
$K_2=n_{p0}^{\gamma-1}K_p/{m_{e0}}{c}^2$.

\begin{figure}[t!]
\centerline{\includegraphics[width=6.8cm]{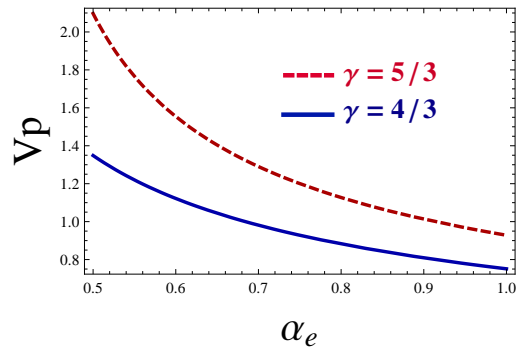}}
\caption{(Color online) Showing the variation of phase speed
$V_p$ with $\alpha_{e}$ for $u_0=0.1$, $\delta=10^{0}$, and
$\alpha_{p}=0.2$. The red dashed line represents the
non-relativistic case and the blue solid line represents the
ultra-relativistic case.} \label{1}
\end{figure}

\begin{figure}[t!]
\centerline{\includegraphics[width=6.8cm]{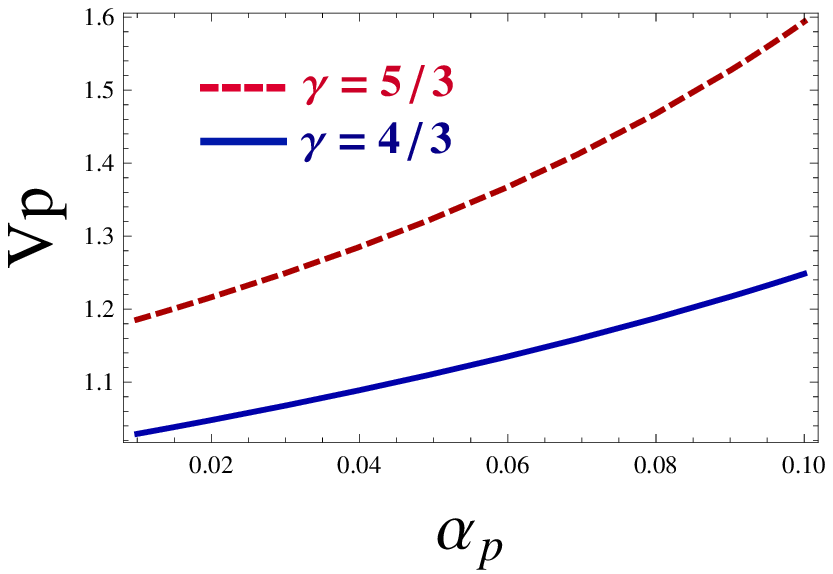}}
\caption{(Color online) Showing the variation of phase speed $V_p$
with $\alpha_{p}$ for $u_0=0.1$, $\delta=10^{0}$, and
$\alpha_{e}=0.4$. The red dashed line represents the
non-relativistic case and the blue solid line represents the
ultra-relativistic case.} \label{2}
\end{figure}

\begin{figure}[t!]
\centerline{\includegraphics[width=6.8cm]{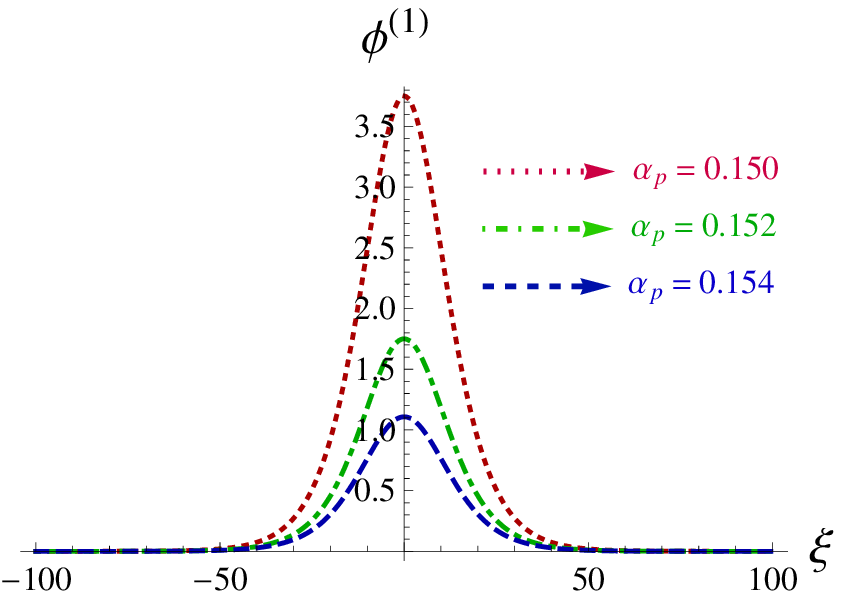}}
\caption{(Color online) Showing the variation of positive
potential K-dV solitons with $\alpha_{p}$ for $u_0=0.1$,
$\alpha_{e}=0.4$, $\omega_{ci}=0.5$, and $\delta=10^{0}$ in case
of non-relativistic limit.} \label{3}
\end{figure}

\begin{figure}[t!]
\centerline{\includegraphics[width=6.8cm]{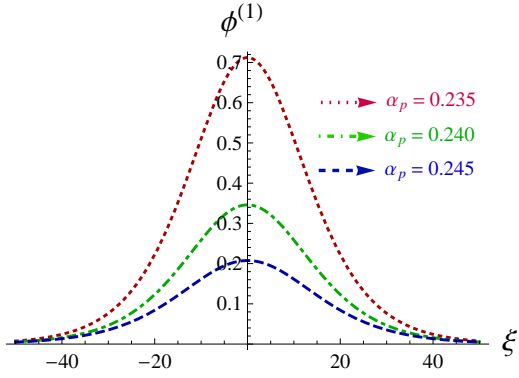}}
\caption{(Color online) Showing the variation of positive
potential K-dV solitons with $\alpha_{p}$ for $u_0=0.1$,
$\alpha_{e}=0.4$, $\omega_{ci}=0.5$, and $\delta=10^{0}$ in case
of ultra-relativistic limit.} \label{4}
\end{figure}

%%%%%%%%%%%%%%%%%%%%%%%%%%%%%%%%%%%%%%%%%%%%%%%%%%%%%%%%%%%%%%%%%%%%%%%%%%%%%%%%%%%%%%%%
\section{K-dV Equation}
To study the dynamics of small but finite amplitude obliquely
propagating NA waves in a magnetized electron-ion (EI) plasma, we
apply the reductive perturbation technique in which independent
variables are stretched as

\begin{eqnarray}
&&\eta=\epsilon^{1/2}(L_xx+L_yy+L_zz-V_pt),
\label{B3a}\\
&&T={\epsilon}^{3/2}t, \label{B3b}
\end{eqnarray}

where $\epsilon$ is a smallness parameter $(0 < \epsilon < 1)$
measuring the amplitude of perturbation, $V_p$ is the wave phase
velocity normalized by the NA speed ($C_i$), and $l_x$, $l_y$,
and $l_z$ are the directional cosines of the wave vector k along
the x, y, and, z axes, respectively, so that $l_x^2$ + $l_y^2$ +
$l_z^2$ = 1. It is noted here that x, y, z are all normalized by
the Debye length $\lambda_{D}$, and $T$ is normalized by the
inverse of light nuclei plasma frequency ($\omega_{pi}^{-1}$ ). We
may expand $n_s$, $u_s$, and $\phi$ in power series of $\epsilon$
as

\begin{eqnarray}
&&n_s=1+\epsilon n_s^{(1)}+\epsilon^{2}n_s^{(2)}+ \cdot \cdot
\cdot, \label{B3c}\\
&&u_{ix,y}=0+\epsilon^{3/2}
u_{ix,y}^{(1)}+\epsilon^{2}u_{ix,y}^{(2)}+\cdot \cdot \cdot,
\label{B3d}\\
&&u_{iz}=0+\epsilon u_{iz}^{(1)}+\epsilon^{2}u_{iz}^{(2)}+\cdot
\cdot \cdot,
\label{B3e}\\
&&\phi=0+\epsilon\phi^{(1)}+\epsilon^{2}\phi^{(2)}+\cdot \cdot
\cdot, \label{B3f}
\end{eqnarray}

now, applying Eqs. (\ref{B3a})-(\ref{B3f}) into Eqs. (\ref{B2a}) -
(\ref{B2e}) and taking the lowest order coefficient of $\epsilon$,
we obtain, $u_{iz}^{(1)}={L_z \phi^{(1)}}/{V_p}$,
 $n_{i}^{(1)}={L_z^2 \phi^{(1)}}/{V_p^2}$,
 $n_{e}^{(1)}=\phi^{(1)}/K_{11}$, $n_{p}^{(1)}=\phi^{(1)}/K_{22}$,
and
$V_p=L_z\sqrt{(\frac{{K_{11}}{K_{22}}}{{K_{22}}{\alpha_e}-{K_{11}}{\alpha_p}})}$
represents the dispersion relation for the NA waves that move
along the propagation vector $k$.

To the lowest order of x- and y-component of the momentum
equation (\ref{B2d}) we get
\begin{eqnarray}
&&u_{iy}^{(1)}=\frac{L_x}{\omega_{ci}}\frac{\partial\phi^{(1)}}{\partial\eta},
\label{B3g}\\
 &&u_{ix}^{(1)}=-\frac{L_y}{\omega_{ci}}\frac{\partial\phi^{(1)}}{\partial\eta}.
 \label{B3h}
\end{eqnarray}
Now, applying Eqs. (\ref{B3a})-(\ref{B3h}) into (\ref{B2d}) one
can obtain from the higher order series of $\epsilon$ of the
momentum and Poisson's equations as

\begin{eqnarray}
 &&u_{iy}^{(2)}=\frac{L_yV_P}{\omega_{ci}^2}\frac{\partial^2\phi^{(1)}}{\partial\eta^2},
 \label{B3i}\\
 &&u_{ix}^{(2)}=\frac{L_xV_P}{\omega_{ci}^2}\frac{\partial^2\phi^{(1)}}{\partial\eta^2},
\label{B3j}\\
&&\frac{\partial^2\phi^{(1)}}{\partial
\eta^2}=n_e^{(2)}\alpha_e-n_i^{(2)}-n_p^{(2)}\alpha_p. \label{B3k}
\end{eqnarray}
We now  follow the same process to get the next higher order
continuity equation as well as z-component of the momentum
equation. Combining these higher order equations together with
Eqs. (\ref{B3g})-(\ref{B3k}) one can obtain the K-dV equation in
the form
\begin{eqnarray}
&&\frac{\partial\phi^{(1)}}{\partial T} + \lambda \phi^{(1)}
\frac{\partial \phi^{(1)}}{\partial \eta}+ \beta \frac{\partial^3
\phi^{(1)}}{\partial \eta^3}=0, \label{B3l}
\end{eqnarray}
where
\begin{eqnarray}
&&\lambda=\frac{V_p^3}{2L_z^2}\left[\frac{{\alpha_e}(\gamma-2)}{K_{11}^2}-\frac{{\alpha_p}(\gamma-2)}{K_{22}^2}-\frac{3L_z^4}{V_p^4}\right],
\label{B4a}\\
&&\beta=\frac{V_p^3}{2L_z^2}\left[1+\frac{(1-L_z^2)}{\omega_{ci}^2}\right].
\label{B4b}
\end{eqnarray}

\begin{figure}[t!]
\centerline{\includegraphics[width=6.8cm]{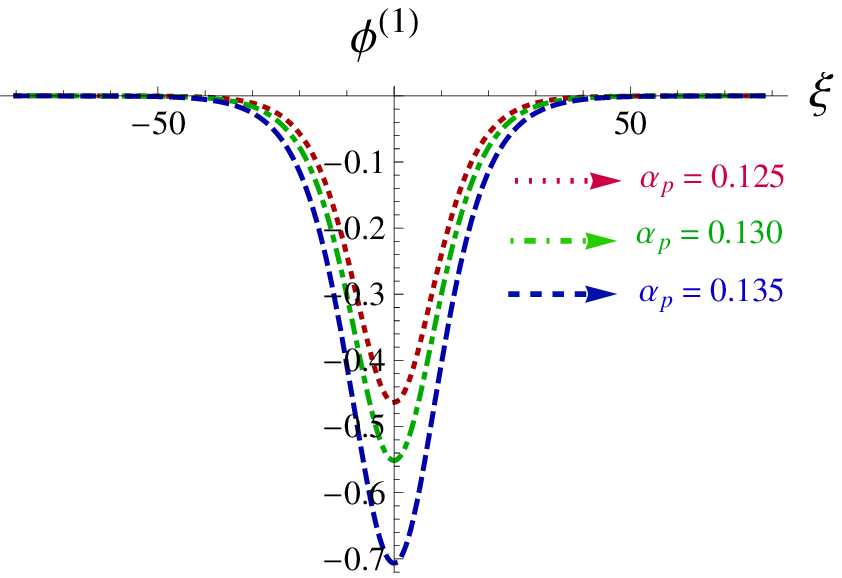}}
\caption{(Color online) Showing the variation of negative
potential K-dV solitons with $\alpha_{p}$ for $u_0=0.1$,
$\alpha_{e}=0.4$, $\omega_{ci}=0.5$, and $\delta=10^{0}$ in case
of non-relativistic limit.} \label{5}
\end{figure}

\begin{figure}[t!]
\centerline{\includegraphics[width=6.8cm]{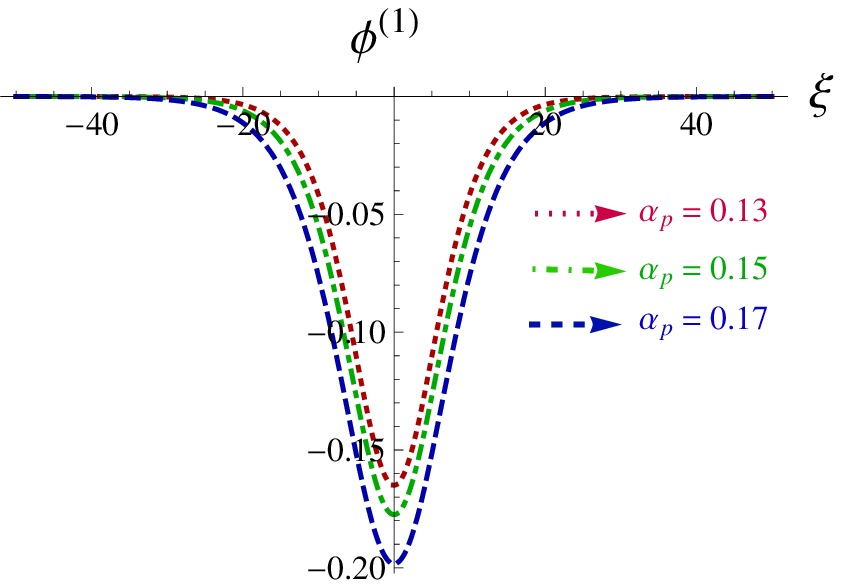}}
\caption{(Color online) Showing the variation of negative
potential K-dV solitons with $\alpha_{p}$ for $u_0=0.1$,
$\alpha_{e}=0.4$, $\omega_{ci}=0.5$, and $\delta=10^{0}$ in case
of ultra-relativistic limit.} \label{6}
\end{figure}

\begin{figure}[t!]
\centerline{\includegraphics[width=6.8cm]{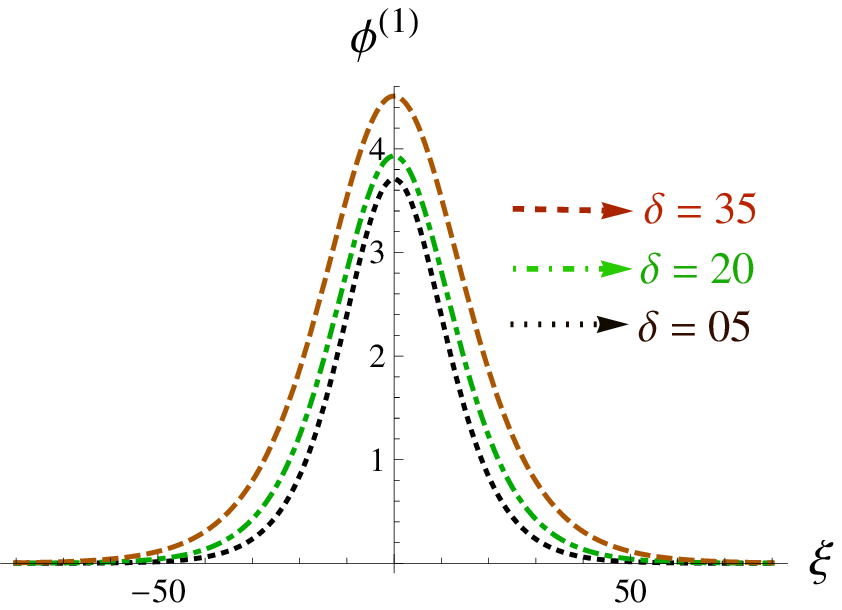}}
\caption{(Color online) Showing the variation of the positive
potential K-dV solitons $\phi^{(1)}$ with the obliqueness of the
wave propagation $\delta$ for $\alpha_{p}>\alpha_{pc}$ and for
$\gamma=5/3$. The other plasma parameters are kept fixed.}
\label{7}
\end{figure}

\begin{figure}[t!]
\centerline{\includegraphics[width=6.8cm]{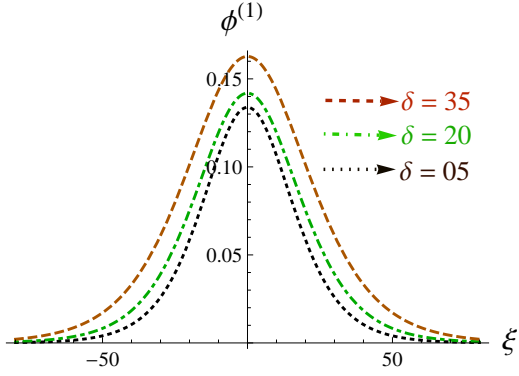}}
\caption{(Color online) Showing the variation of the positive
potential K-dV solitons $\phi^{(1)}$ with the obliqueness of the
wave propagation $\delta$ for $\alpha_{p}>\alpha_{pc}$ and for
$\gamma=4/3$. The other plasma parameters are kept fixed.}
\label{8}
\end{figure}

To observe the influence of different plasma parameters on the
propagation of solitary waves in magnetized quantum plasma, we
derive the solution of K-dV equation (\ref{B3l}). The stationary
solitary wave solution of standard K-dV equation is obtained by
considering a frame $\xi=\eta-u_{0}T$ (moving with speed $u_{0}$)
and the solution is,
\begin{eqnarray}
{\rm \phi^{(1)}}=\rm \phi_m{\rm[sech^{2}}(\frac{\xi}{\Delta})],
\label{solK-dV}
\end{eqnarray}
where the amplitude, $\phi_m=3u_{0}/\lambda$, and the width,
$\Delta=(4\beta/u_{0})^{1/2}$

\section{MK-dV Equation}
To obtain the MK-dV equation, same stretched co-ordinates is
applied as we used in K-dV equation in section III (i.e.,
Eqs.(\ref{B3a}) and (\ref{B3b})) and also used the dependent
variables which is expanded as
\begin{eqnarray}
&&\hspace*{-10mm}n_s=1+\epsilon^{1/2} n_s^{(1)}+\epsilon
n_s^{(2)}+\epsilon^{3/2} n_s^{(3)} +\cdot \cdot
\cdot, \label{B4c}\\
&&\hspace*{-10mm}u_{ix,y}=0+\epsilon
u_{ix,y}^{(1)}+\epsilon^{3/2}u_{ix,y}^{(2)}+\epsilon^{2}u_{ix,y}^{(3)}+\cdot
\cdot \cdot,
\label{B4d}\\
&&\hspace*{-10mm}u_{iz}=0+\epsilon^{1/2} u_{iz}^{(1)}+\epsilon
u_{iz}^{(2)}+\epsilon^{3/2} u_{iz}^{(3)}+\cdot \cdot \cdot,
\label{B4e}\\
&&\hspace*{-10mm}\phi=0+\epsilon^{1/2}\phi^{(1)}+\epsilon\phi^{(2)}+\epsilon^{3/2}\phi^{(3)}+\cdot
\cdot \cdot, \label{B4f}
\end{eqnarray}
we find the same expressions of $n_{i}^{(1)}$, $n_{e}^{(1)}$,
$n_{p}^{(1)}$, $u_{iz}^{(1)}$, $u_{ix,y}^{(1)}$, $u_{ix,y}^{(2)}$
and $V_p$ by using the values of $\eta$ and $T$ in
Eqs.(\ref{B2a})-(\ref{B2e}) and (\ref{B4c})-(\ref{B4f}) as before
in section III. The next higher order series of $\epsilon$ of
continuity, momentum and poisson's equations as
\begin{eqnarray}
&&\hspace*{-12mm}\frac{\partial n_i^{(1)}}{\partial
T}-V_p\frac{\partial n_i^{(3)}}{\partial\eta}+L_x\frac{\partial
u_{ix}^{(2)}}{\partial\eta}+L_x\frac{\partial}{\partial\eta}(n_i^{(1)}u_{ix}^{(1)})\nonumber\\
&&\hspace*{-7mm}+L_y\frac{\partial
u_{iy}^{(2)}}{\partial\eta}+L_y\frac{\partial}{\partial\eta}(n_i^{(1)}u_{iy}^{(1)})+L_z\frac{\partial
u_{iz}^{(3)}}{\partial\eta}\nonumber\\
&&\hspace*{-7mm}+L_z\frac{\partial}{\partial\eta}(n_i^{(1)}u_{iz}^{(2)})+L_z\frac{\partial}{\partial\eta}(n_i^{(2)}u_{iz}^{(1)})=0,
\label{B4g}\\
&&\hspace*{-12mm}\frac{\partial u_{iz}^{(1)}}{\partial
T}-V_p\frac{\partial
u_{iz}^{(3)}}{\partial\eta}+L_z\frac{\partial}{\partial\eta}(u_{iz}^{(1)}u_{iz}^{(2)})+L_z\frac{\partial\phi^{(3)}}{\partial\eta}=0,
\label{B4h}\\
&&\hspace*{-12mm}L_z\frac{\partial\phi^{(3)}}{\partial\eta}-K_{11}L_z\frac{\partial
n_e^{(3)}}{\partial\eta}-F\frac{\partial}{\partial\eta}(n_e^{(1)}n_e^{(2)})=0,
\label{B4i}\\
&&\hspace*{-12mm}L_z\frac{\partial\phi^{(3)}}{\partial\eta}-K_{22}L_z\frac{\partial
n_p^{(3)}}{\partial\eta}-G\frac{\partial}{\partial\eta}(n_p^{(1)}n_p^{(2)})=0,
\label{B4j}\\
&&\hspace*{-12mm}\frac{\partial^2\phi^{(1)}}{\partial\eta^2}=\alpha_en_e^{(3)}-n_i^{(3)}-\alpha_pn_p^{(3)},
\label{B4k}
\end{eqnarray}

where $F=K_{11}L_z(\gamma-2)$ and $G=K_{22}L_z(\gamma-2)$.

Now combining these higher order equations together with
Eqs.(\ref{B4g})-(\ref{B4k}) one can obtain
\begin{eqnarray}
&&\hspace*{-10mm}\frac{\partial\phi^{(1)}}{\partial T} +
M\phi^{(1)2} \frac{\partial \phi^{(1)}}{\partial
\eta}+N\frac{\partial^3 \phi^{(1)}}{\partial \eta^3}=0.
\label{DIAK-dV}
\end{eqnarray}
This is well-known MK-dV equation that describes the obliquely
propagating NA waves in a magnetized quantum plasma. where $M$
and $N$ are given by
\begin{eqnarray}
&&\hspace*{-10mm}M=\frac{V_p^3}{2L_z^2}\left[\frac{(\gamma-2)^2\alpha_p}{9K_{22}^3}-\frac{(\gamma-2)^2\alpha_e}{9K_{11}^3}+\frac{5L_z^6}{6V_p^6}\right],
\label{B5a}\\
&&\hspace*{-10mm}N=\frac{V_p^3}{2L_z^2}\left[1+\frac{(1-L_z^2)}{\omega_{ci}^2}\right].
\label{B5b}
\end{eqnarray}

\begin{figure}[t!]
\centerline{\includegraphics[width=6.8cm]{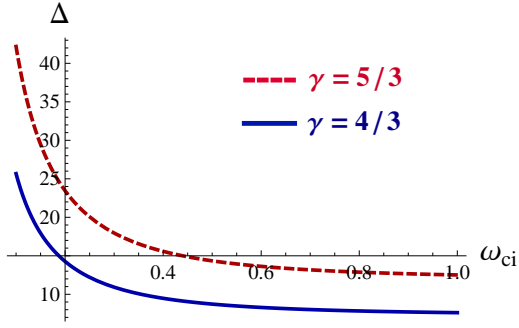}}
\caption{(Color online) Variation of $\Delta$ on solitary profile
for different values of $\omega_{ci}$ when electron and positron
being non-relativistic and ultra-relativistic degenerate.}
\label{11}
\end{figure}

\begin{figure}[t!]
\centerline{\includegraphics[width=6.8cm]{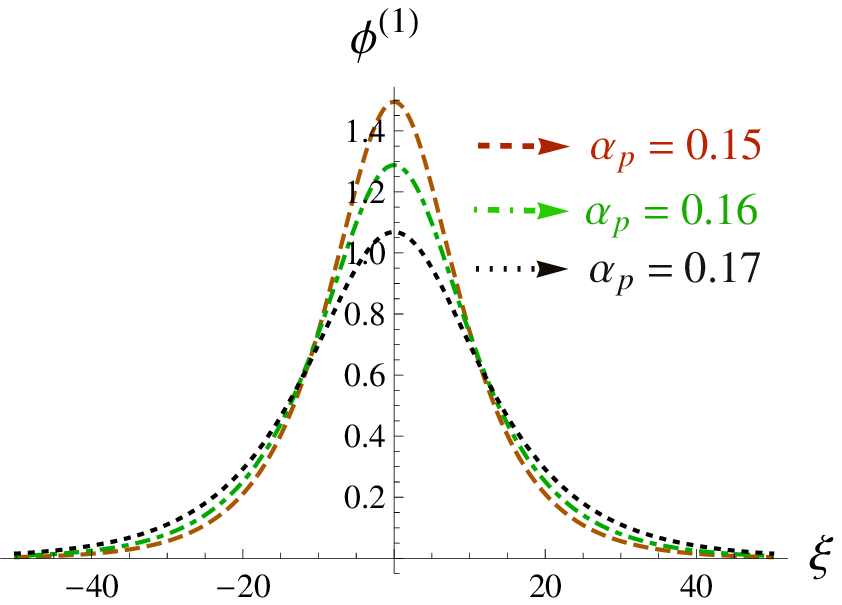}}
\caption{(Color online) Showing the variation of the amplitude of
magnetized MK-dV solitons with $\alpha_p$ for
$\alpha_p>\alpha_{pc}$ (in case of non-relativistic degenerate).
The other plasma parameters are kept fixed.} \label{12}
\end{figure}

\begin{figure}[t!]
\centerline{\includegraphics[width=6.8cm]{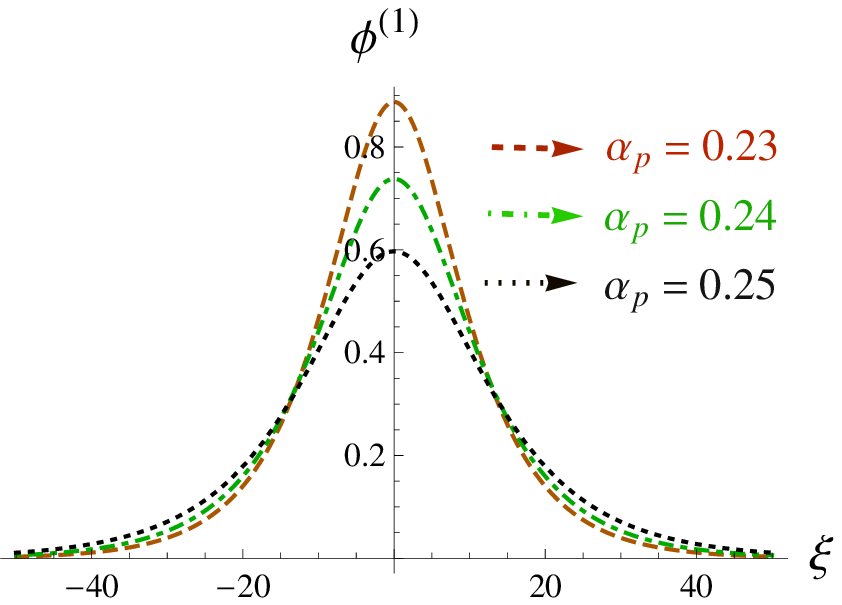}}
\caption{(Color online) Showing the variation of the amplitude of
magnetized MK-dV solitons with $\alpha_p$ for
$\alpha_p>\alpha_{pc}$ (in case of ultra-relativistic
degenerate). The other plasma parameters are kept fixed.}
\label{13}
\end{figure}

\begin{figure}[t!]
\centerline{\includegraphics[width=6.8cm]{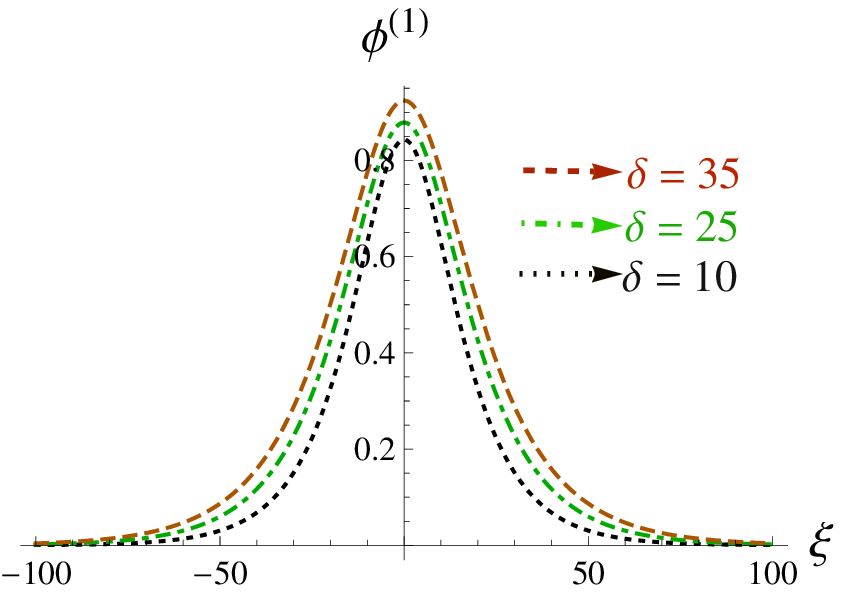}}
\caption{(Color online) Showing the variation of the amplitude of
magnetized MK-dV solitons with $\delta$ for
$\alpha_p>\alpha_{pc}$ (in case of non-relativistic degenerate).
The other plasma parameters are kept fixed.} \label{14}
\end{figure}

\begin{figure}[t!]
\centerline{\includegraphics[width=6.8cm]{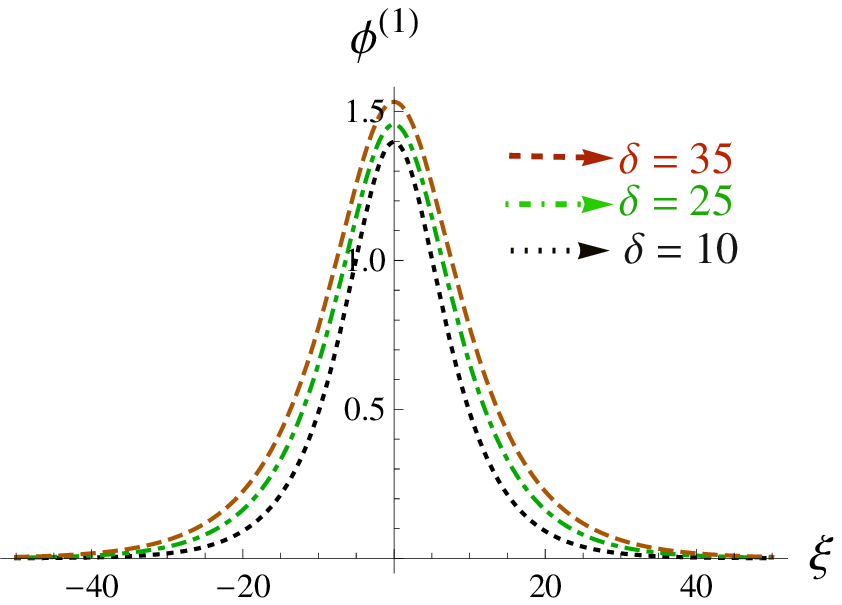}}
\caption{(Color online) Showing the variation of the amplitude of
magnetized MK-dV solitons with $\delta$ for
$\alpha_p>\alpha_{pc}$ (in case of ultra-relativistic
degenerate). The other plasma parameters are kept fixed.}
\label{15}
\end{figure}
The stationary solitary wave solution of standard MK-dV equation
is obtained by considering a frame $\xi=\eta-u_{0}T$ (moving with
speed $u_{0}$) and the solution is,
\begin{eqnarray}
{\rm \phi^{(1)}}=\rm \phi_m{\rm[sech}(\frac{\xi}{\varpi})],
\label{solK-dV}
\end{eqnarray}
where the amplitude, $\phi_m=\sqrt{(6u_{0}/M)}$ and the width
$\varpi= {\sqrt{(N/u_{0})}}$.
%%%%%%%%%%%%%%%%%%%%%%%%%%%%%%%%%%%%%%%%%%%%%%%%%%%%%%%%%%%%%%%%%%%%%%%%%%%%%%
\section{SG Equation}
To drive the SG equation for NA SWs, we utitize the second order
equation of surface charge density $\rho$ by analyzing Eqs.
(1)-(5) and solution of which provides $\lambda=0$ since
$\phi^{(1)}=0$. It is noted that at $\lambda=0$ its critical value
$\alpha_p=\alpha_{pc}$. For $\alpha_p$ around its critical values
$(\alpha_{pc})$, $\lambda=\lambda_0$ can be expressed as
\begin{eqnarray}
&&\lambda_0\simeq s\left(\frac{\partial \lambda}{\partial
\alpha_p}\right)_{\alpha_p=\alpha_{pc}}|\alpha_p-\alpha_{pc}|=sC_{1}\epsilon^{1/2},
\label{B5c}
\end{eqnarray}
where $|\alpha_p-\alpha_{pc}|$ is a small and dimensionless
parameter, and can be taken as the expansion parameter
$\epsilon^{1/2} $, i.e. $|\alpha_p-\alpha_{pc}|\simeq
\epsilon^{1/2} $, and $s=1$ for $\alpha_p<\alpha_{pc}$ and $s=-1$
for $\alpha_p>\alpha_{pc}$. $C_{1}$ is a constant which is given
by
\begin{eqnarray}
C_{1}=-\frac{(\gamma-2)}{K_{22}^2}.\label{B5d}
\end{eqnarray}
The surface charge density $\rho^{(2)}$ can be expressed as
\begin{eqnarray}
&&\epsilon\rho^{(2)}\simeq
-\frac{1}{2}\epsilon^{3/2}C_{1}s({\phi^{(1)}})^2, \label{B5e}
\end{eqnarray}
which, therefore, must be included in the third order Poisson's
equation. To the next higher order in $\epsilon^{3/2}$, we obtain
the following equation
\begin{eqnarray}
&&\hspace{-15mm}\frac{\partial^2\phi^{(1)}}{\partial
\eta^2}+\frac{1}{2}sC_{1}({\phi^{(1)}})^2-\alpha_en_{e}^{(3)}+n_{e}^{(3)}\nonumber\\
&&\hspace{-10mm}+\alpha_p[\frac{1}{K_{22}}\frac{\partial\phi^{(3)}}{\partial\eta}-
\frac{(\gamma-2)}{K_{22}^2}\frac{\partial}{\partial\eta}(\phi^{(1)}\phi^{(2)})\nonumber\\
&&\hspace{-10mm}+\frac{(\gamma-2)^2}{3K_{22}^2}\frac{\partial(\phi^{(1)})^3}{\partial\eta}]=0.\label{B5f}
\end{eqnarray}
After simplification, we can write from Eq. (\ref{B5f})
\begin{eqnarray}
\frac{\partial\phi^{(1)}}{\partial
T}+sC_{1}\alpha\phi^{(1)}\frac{\partial \phi^{(1)}}{\partial
\eta} +\alpha\beta({\phi^{(1)}})^2\frac{\partial
\phi^{(1)}}{\partial \eta}\nonumber\\
&&\hspace*{-50mm} +\alpha\gamma\frac{\partial^3
\phi^{(1)}}{\partial \eta^3}=0. \label{sGS}
\end{eqnarray}
Equation (\ref{sGS}) is known as SG equation. It supports both
the SWs and DLs solutions since it contains both $\phi$-term of
K-dV and $\phi^2$-term of mK-dV equation. The Gardner equation
derived here is valid for $\alpha_p\simeq \alpha_{pc}$.

\begin{figure}[t!]
\centerline{\includegraphics[width=6.8cm]{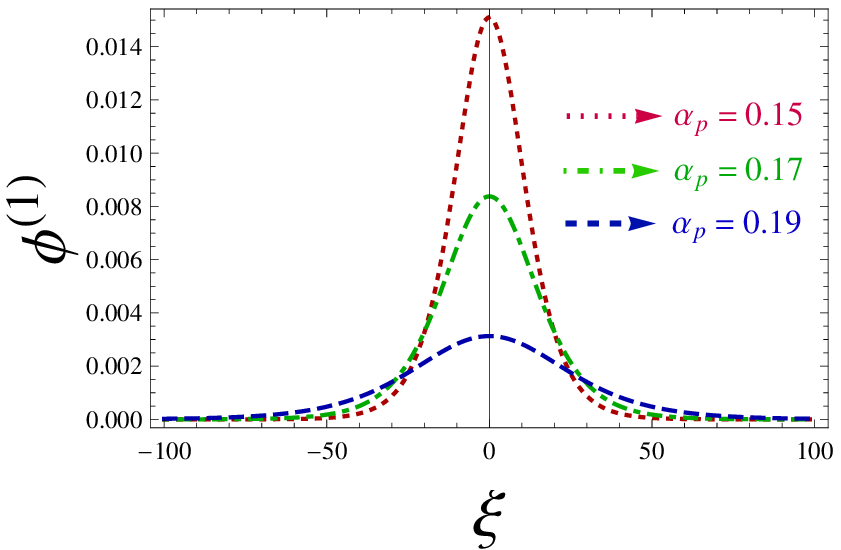}}
\caption{(Color online) Showing the variation of the amplitude of
positive magnetized SGs solitons with $\alpha_p$ for
$\alpha_p>\alpha_{pc}$ (in case of non-relativistic degenerate).
The other plasma parameters are kept fixed.} \label{16}
\end{figure}
\begin{figure}[t!]
\centerline{\includegraphics[width=6.8cm]{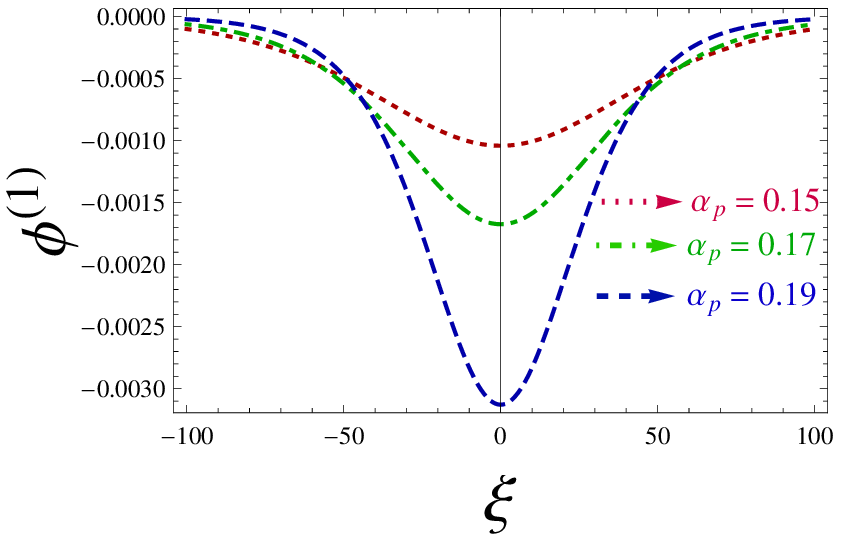}}
\caption{(Color online) Showing the variation of the amplitude of
negative magnetized SGs solitons with $\alpha_p$ for
$\alpha_p>\alpha_{pc}$ (in case of non-relativistic degenerate).
The other plasma parameters are kept fixed.} \label{17}
\end{figure}
\begin{figure}[t!]
\centerline{\includegraphics[width=6.8cm]{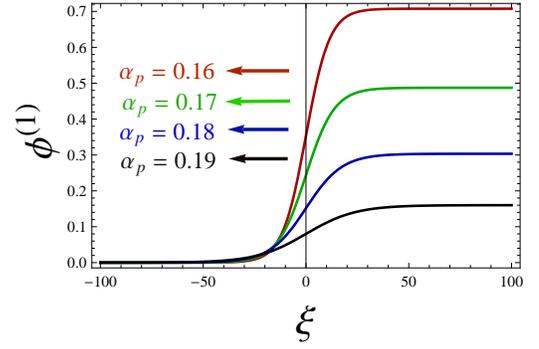}}
\caption{(Color online) Showing the variation positive potential
DLs solitons with $\alpha_p$ in case of non-relativistic
degenerate. The other plasma parameters are kept fixed.}
\label{18}
\end{figure}
%%%%%%%%%%%%%%%%%%%%%%%%%%%%%%%%%%%%%%%%%%%%%%%%%%%%%%%%%%%%%%%%%%%%%%%%%%%%%%%%%%%%%%%%%%%%%%%%%%%%%%%%%%%%%%%%%%%%%%%%%%%%%%%%%%%%%%%%%%%%%%%%%%%%%%%%%%%%%%%

By applying different boundary conditions in Eq. (38) one can
find two type of solutions, namely solitary wave solution and
double layer solution which are common in plasma literature
\cite{Mannan2012,Alam2014}.

The SW solution of SG equation is given by the following equation:
\begin{eqnarray}
\phi^{(1)}=\left[\frac{1}{\phi_{m2}}-\left(\frac{1}{\phi_{m2}}
-\frac{1}{\phi_{m1}}\right)\cosh^2\left(\frac{\xi}{\Lambda}\right)\right]^{-1},
\label{B5g}
\end{eqnarray}
where
\begin{eqnarray}
&&\phi_{m1,2}=\phi_m\left[1\mp\sqrt{1+\frac{U_0}{V_0}}\right],\label{B6a}\\
&&\hspace*{-5mm}U_0=\frac{C_1s\alpha}{3}\phi_{m1,2}+\frac{\alpha\beta}{6}\phi^2_{m1,2},\label{B6b}\\
&&V_0=\frac{C_{1}^2s^2\alpha}{6\beta},\label{B6c}\\
&&\phi_m=\frac{-C_{1}s}{\alpha},\label{B6d}\\
&&\Lambda=\frac{2}{\sqrt{-\gamma\phi_{m1}\phi_{m2}}},\label{B6e}\\
&&\gamma=\frac{\beta}{6}.\label{B6f}
\end{eqnarray}

The DL solution of Eq. (\ref{sGS}) is given by
\begin{eqnarray}
\phi^{(1)}=\frac{\phi_m}{2}\left[1+\tanh \left
(\frac{\xi}{\Lambda} \right)\right],\label{SolDL}
\end{eqnarray}
with
\begin{eqnarray}
U_0=-\frac{s^2\alpha}{6\beta},\label{B6g}\\
\phi_m=\frac{6U_0}{s\alpha},\label{B6h}\\
\Lambda=\frac{2}{\phi_m\sqrt{-\gamma}}, \label{B6i}
\end{eqnarray}

where $\gamma=\beta/6$ and $\phi_m$ ($\Lambda$) is the DL height
(thickness).
%%%%%%%%%%%%%%%%%%%%%%%%%%%%%%%%%%%%%%%%%%%%%%%%%%%%%%%%%%%%%%%%%%%%%%%%%%%%%.
\section{Discussions}
We considered a relativistic magnetized four component EPI plasma
system containing non-degenerate inertial light nuclei, both
non-relativistically and ultra-relativistically degenerate
electrons and positrons, and immobile heavy nuclei. The
well-known reductive perturbation method has been used to drive
the K-dV equation, and then solve it to examine an exact
analytical expression for small amplitude solitary waves. In
order to trace the effect of higher order nonlinearity, the mK-dV
and SG equations have been derived since K-dV equation is valid
only for the limits ¸ $\lambda\neq0$, ¸ $\lambda>0$ and¸
$\lambda<0$. By using this method, the K-dV, MK-dV and Gardner
equations for NA SWs for ultradensed magnetized EPI plasmas are
obtained and numerically analyzed the different plasma
parameters. We have observed and analyzed that both compressive
and rarefactive solitary waves (SWs) could exist for K-dV and
GSs, but only rarefactive structures can be formed for MK-dV and
DLs structures. We have observed K-dV solitons of both positive
(Figs 3, 4, 7 and 8) and negative (Figs 5 and 6) potentials but
that of positive (Figs 10, 11, 12 and 13 ) only for MK-dV
solitons. On proceeding to more higher order calculation, we
obtained Gardner equation as well as its solution, and from the
GSs shown in Figs 14 and 15, we have investigated that both
positive and negative potential GSs can be observed in the
system. Further, we have studied and analyzed the DLs solution,
and observed that only positive potential DL structure (Fig 16)
can be found in the system. The plasma system under consideration
supports finite amplitude solitary, gardner and double layer
structures and their solutions are stable. The amplitude of SWs,
GSs and DLs has been modified by the effect of degenerate
pressure of electrons and positrons, this degenerate pressure is
illustrated from the non-relativistic $({P_e}\propto{n_e}^{5/3})$
to ultra-relativistic $({P_e}\propto{n_e}^{4/3})$ regime. By
using Chandrasekhar's equation of state for relativistically
degenerate electrons and positrons, it is examined that the
relativistic factor greatly affects the speed of NA SWs where for
the non-relativistic degenerate electrons and positrons,
$\gamma=\frac{5}{3}$ and for the ultra-relativistic degenerate
electrons and positrons, $\gamma=\frac{4}{3}$, and thus found
that the relativistic factor, $\gamma=5/3 ~(non-relativistic)
> \gamma=4/3 ~(ultra-relativistic)$ in every cases. In all
cases, it is noticed that the relativistic corrections decrease
the wave frequency, which, in the ultra-relativistic limit is
smaller than the corresponding non-relativistic limit. The
numerical results that have been found from our investigation are
plotted in Figs. 1 - 16 and can be summarized as follows:

\begin{enumerate}
\item{The four component EPI plasma system under consideration supports finite amplitude NA SWs, whose fundamental
features (viz., amplitude, speed, and width) strongly depend on
different plasma parameters, particularly, positron to ion number
density ratio $\alpha_p$, obliqueness, cyclotron frequency
$\omega_{ci}$, and the relativistic factor (i.e.,
non-relativistic limit and ultra-relativistic limit). We have
numerically obtained here that for $\lambda=0$, the amplitude of
the K-dV solitons become infinitely large, and the K-dV solution
is no longer valid at $\lambda\simeq0$. It has been observed that
the solution of the K-dV equation supports both compressive
(positive) and rarefactive (negative) structures depending on the
critical value of $\alpha_p$.}

\item{In our present investigation, we have found that for
$\alpha_{pc}=0.148$, the amplitude of the SWs breaks down due to
the vanishing of the nonlinear coefficient $\lambda$. We have
observed that at $\alpha_p>0.148$, positive (compressive)
potential SWs exist, whereas at $\alpha_p<0.148$, negative
(rarefactive) SWs exist (shown in Figs. 3-8).}

\item{It is found that the phase speed of the NA SWs decreases with
the increasing values of electron to light nuclei number density
ratio $\alpha_e$ and increases as the increasing values of
positron to light nuclei number density ratio $\alpha_p$. The
phase speed is always higher for the non-relativistic case than
the ultra-relativistic case shown in Figs. 1 and 2.}

\item{It is shown that with the increase of $\alpha_p$, the
amplitude and width of solitons waning gradually for positive
potential, while the width of the solitons swelling gradually for
negative potential. It is also seen that for non-relativistic
limit the amplitude of the solitary waves is found to be greater
than the ultra-relativistic limit shown in Figs. 3,4,5 and 6.}

\item{The variation of the amplitude of NA SWs is taken place with the different values of obliqueness of the wave
propagation. It is observed that with the increase in obliqueness
of the wave propagation, there is a increase in amplitude of the
magnetized K-dV solitons which is exhibited in Figs. 7 and 8. It
is noticed that as the value of $\delta$ increases, the amplitude
of the solitary waves increases, while their width increases for
the lower range of $\delta$ (from $0^{\circ}$ to about
$50^{\circ}$), and decreases for its higher range (from
$50^{\circ}$ to about $90^{\circ}$). We note that as $\delta
\rightarrow90^{\circ}$, the width goes to $0$, and the amplitude
goes to $\infty$. This means that in case of larger values of
$\delta$, the wave amplitude becomes large enough to break down
the validity of the reductive perturbation method
\cite{Alinejad2012}. Our present investigation is only valid for
small value of $\delta$ but invalid for arbitrarily large value
of $\delta$.} The results that we have investigated in our present
investigation support the results of Haider et al.
\cite{Haider2012a,Haider2012b}.

\item{Figure 9 shows that the width of the K-dV solitary profiles decrease with the
increasing values of $\omega_{ci}$ for both non-relativistic and
ultra-relativistic case. The solitary waves is also found to be
greater for non-relativistic case than the ultra-relativistic
case.}

\item{For MK-dV solitons, only compressive solitons are obtained, which is common in literature \cite{Rahman2014}. It
is observed that the apmlitude of MK-dV solitons decreases with
the increasing of $\alpha_p$ like as K-dV soliton as shown in
Fig. 10 for non-relativistic limit and Fig. 11 for
ultra-relativistic limit.}

\item{The amplitude of the magnetized MK-dV solitons affects notably by the obliqueness effect. From Figs. 12 and 13, we found that the amplitude of
MK-dV solitons increases with the increasing of obliqueness.}

\item{In case of GSs, both the positive and negative structures are
formed. It is investigated that like K-dV and MK-dV soliton the
amplitude of magnetized GSs are decreases with the increasing of
$\alpha_p$ as shown in Fig. 14 for positive potential and Fig. 15
for negative potential in non-relativistic limit.}

\item{It is observed that only positive potential DLs are exist. The
amplitude of positive potential DLs decrease with the increasing
of $\alpha_p$ as shown in Fig. 16.}

\end{enumerate}
We note that we have investigated obliquely propagating
one-dimensional solitary structures by deriving the K-dV and
MK-dV equations. However, one can derive Zakarov-Kuznetsove (ZK)
equation  to study multi-dimensional solitary structures and
their multi-dimensional instability \cite{Mamun1988}, which is
beyond the scope of our present work. It emphasizes here that the
findings of our present investigation should be useful for
understanding the basic features of the obliquely propagating NA
SWs in ultra-relativistic degenerate magnetized plasmas which
occur in many astrophysical compact objects, like white dwarfs,
neutron stars, magnetars, etc.

\section{Acknowledgments}

B. Hosen, M. G. Shah and M. R. Hossen would like to thank
Ministry of Science and Technology (MOST), Government of
Bangladesh, for awarding the National Science and Technology
(NST) fellowship.
%%%%%%%%%%%%%%%%%%%%%%%%%%%%%%%%%%%%%%%%%%%%%%%%%%%%%%%%%%%%%%%%%%%%%%%%%%%%%%%%%%%%%%%%%%%%%%%%%%%%%%%%%%%%%%%%%%%%%%%%%%%%%%%%%%%%%%%%

\end{document}